\documentclass[aps,prb,twocolumn]{revtex4} 

\usepackage{graphicx}
\usepackage{dcolumn}
\usepackage{bm}

\newcommand{\comment}[1]{}

\begin{document}

\title{Simplified Derivation of the Non-Equilibrium Probability Distribution}


\author{Phil Attard}

\date{\today. phil.attard1@gmail.com}

\begin{abstract}
A simple and transparent derivation
of the formally exact probability distribution
for classical non-equilibrium systems is given.
The corresponding stochastic, dissipative equations of motion
are also derived.
\end{abstract}

\pacs{}

\maketitle

                \section*{Introduction}

Since the beginning,
the Maxwell-Boltzmann probability distribution
has been the center piece of equilibrium statistical mechanics.
\cite{Feynman98,Pathria72,McQuarrie00,TDSM}
In contrast,
even today there is no consensus
for the corresponding probability distribution
for non-equilibrium systems.
\cite{Kubo78,Zwanzig01,Bellac04,Pottier10,NETDSM}

The most common approximation
for a non-equilibrium system that has a time-varying potential
is the Maxwell-Boltzmann distribution
evaluated at the current time.
The problem with this is that it is insensitive to the sign
of the molecular velocities,
which violates the second law of thermodynamics.
Another approximation
is the so-called Yamada-Kawasaki distribution,
which also uses the Maxwell-Boltzmann distribution,
but evaluated at point on the adiabatic (Hamiltonian) trajectory
in the distant past.\cite{Yamada67,Yamada75}
This takes the non-equilibrium system
to be instantaneously in equilibrium in the past,
again in violation of the second law,
and it neglects the influence of the reservoir
on the subsequent evolution of the sub-system.
Neither approximation has been generalized
to thermodynamic non-equilibrium systems
(heat flow, chemical reactions, etc.).

The present author has derived a formally exact expression
for the non-equilibrium  probability distribution
that is suitable for systems with time varying potentials
or with  applied thermodynamic gradients.\cite{AttardV}
Although still lacking broad consensus,
it is true to say that it is the only
non-equilibrium probability distribution that
respects the second law of thermodynamics
and that has been successfully tested analytically
and numerically with computer simulation.
\cite{AttardV,AttardVIII,AttardIX,Attard09}
The most complete account of the theory and the various tests
is given in Ref.~[\onlinecite{NETDSM}].
The result has recently been generalized to the non-equilibrium quantum case.
\cite{QSM4}

This paper presents a simplified and shortened derivation
of this non-equilibrium probability distribution,
replacing some of the mathematical derivation of  Ref.~[\onlinecite{NETDSM}]
by physical arguments.
Some clarifying remarks are added in places,
and some approximate steps in the original derivation
are either removed or corrected.

                \section{Reservoir Entropy} \label{Sec:NE-prob}

\subsection{Trajectory Entropy} \label{Sec:TrajEnt}

The first task is to obtain the reservoir entropy
for a trajectory in the sub-system phase space.
Because points in phase space have no internal entropy
(i.e.\ they have uniform weight),
the points in phase space of the sub-system itself have no entropy,
and so the reservoir entropy is the same as the total entropy.\cite{TDSM}

For a mechanical non-equilibrium system
the time dependent Hamiltonian has the form
\begin{equation}
{\cal H}({\bf \Gamma},t)
=
{\cal H}^\mathrm{bare}({\bf \Gamma})
+ U^\mathrm{ext}({\bf \Gamma}_\mathrm{q},t) .
\end{equation}
Here ${\bf \Gamma} $
is a point in the phase space point of the sub-system,
which may be split into position  and momentum components,
${\bf \Gamma} = \{{\bf \Gamma}_\mathrm{q}, {\bf \Gamma}_\mathrm{p}\}$.
The total energy is not fixed,
but depends upon the work done.
The latter  depends upon the specific trajectory
leading up to the present point,
and it is given by
\begin{eqnarray} \label{Eq:WorkulG}
W(\underline {\bf \Gamma} )
& = &
\int_{0}^t \mathrm{d} t'\,
\frac{\partial U^\mathrm{ext}({\bf \Gamma}_\mathrm{q}(t'),t') }{\partial t'}
\nonumber \\ & = &
\int_{0}^t \mathrm{d} t'\,
\dot {\cal H}^0({\bf \Gamma},t') .
\end{eqnarray}
The the change in total energy of the system
is the work done,
$\Delta E_\mathrm{tot}(\underline {\bf \Gamma} )
\equiv W(\underline {\bf \Gamma} )$.
The adiabatic rate of change of the sub-system energy is
\begin{equation} \label{Eq:dotH0}
\dot {\cal H}^0({\bf \Gamma},t)
\equiv
\frac{\partial {\cal H}({\bf \Gamma},t) }{\partial t}
+ \dot{\bf \Gamma}^0 \cdot \nabla {\cal H}({\bf \Gamma},t)
=
\frac{\partial U^\mathrm{ext}({\bf \Gamma}_\mathrm{q},t) }{\partial t} .
\end{equation}

The sub-system has an energy that only depends upon the current point and time.
Hence the change in the sub-system energy over a trajectory
is just the difference between the initial and final values
of the Hamiltonian,
\begin{equation}
\Delta E_\mathrm{s}(\underline {\bf \Gamma})
=
{\cal H}({\bf \Gamma},t) - {\cal H}({\bf \Gamma}_0,0),
\end{equation}
where the trajectory $\underline {\bf \Gamma}$
starts at ${\bf \Gamma}_0$ at time $t=0$
and ends at ${\bf \Gamma}$ at time $t$.

With the change in reservoir energy being the change in total
energy less the change in sub-system energy
one can write the change in reservoir entropy
on  a particular trajectory as
\begin{eqnarray} \label{Eq:DStraj}
S_\mathrm{r}(\underline {\bf \Gamma})
& = &
\frac{\Delta E_\mathrm{r}(\underline {\bf \Gamma})}{T}
\nonumber \\ & = &
\frac{W(\underline {\bf \Gamma})
- \Delta E_\mathrm{s}(\underline {\bf \Gamma}) }{T}
\nonumber \\ & = &
\frac{ - {\cal H}({\bf \Gamma},t) }{T}
+ \frac{1}{T} \int_{0}^t \mathrm{d} t'\,
\dot {\cal H}^0({\bf \Gamma}(t'),t')
\nonumber \\ & \equiv &
S_\mathrm{st}({\bf \Gamma},t)
+
S_\mathrm{dyn}(\underline {\bf \Gamma}) ,
\end{eqnarray}
where $T$ is the reservoir temperature.
This is strictly the change from the initial time,
but the initial values,
including the initial sub-system energy
${\cal H}({\bf \Gamma}(0),0)$ in the third equality,
are not shown.
In the final equality
has been identified the static reservoir entropy
for a mechanical non-equilibrium system,
\begin{equation} \label{Eq:SreqMechWk}
S_\mathrm{st}({\bf \Gamma},t)
=
\frac{ - {\cal H}({\bf \Gamma},t) }{T} .
\end{equation}
More generally,
this is obtained from the exchange of conserved quantities
with the reservoir
and is the usual equilibrium formula for the entropy
(see, for example, Eq.~(\ref{Eq:Sst-E0-E1})).
The dynamic part of the reservoir entropy,
which is the specifically non-equilibrium part of the reservoir entropy,
is defined as
\begin{eqnarray} \label{Eq:SrneMechWk}
S_\mathrm{dyn}(\underline {\bf \Gamma})
& \equiv &
-\int_{0}^t \mathrm{d} t'\,
\dot S_\mathrm{st}^0({\bf \Gamma}(t'),t')
\nonumber \\ & = &
\frac{1}{T} \int_{0}^t \mathrm{d} t' \,
\dot {\cal H}^0({\bf \Gamma}(t'),t') .
\end{eqnarray}
This term vanishes for an equilibrium system.
It represents a correction for double counting in the expression for
the static entropy when it is applied to a non-equilibrium system.

The double counting may be seen by noting that
$S_\mathrm{st}({\bf \Gamma},t)$ comes from
the total change in the sub-system energy.
This total change is in part due to the adiabatic evolution of the sub-system,
which is independent of the reservoir,
and in part due to interactions with the reservoir.
By energy conservation, only the latter change the reservoir entropy.
This is why the adiabatic changes of the sub-system
have to be subtracted
from the total  change in  reservoir entropy
via $S_\mathrm{dyn}(\underline {\bf \Gamma})$.

\subsection{Reduction to the Point Entropy} \label{Sec:PtEnt}

Denote the most likely trajectory
that ends at  ${\bf \Gamma}$ at time $t$ by an over-line,
\begin{equation}
\overline{ \underline {\bf \Gamma}}
\equiv
\overline{ {\bf \Gamma}}(t'|{\bf \Gamma},t),
\;\;  t' \le t .
\end{equation}

The reduction theorem\cite{AttardV} states that
the entropy of the current point
is equal to the entropy of the most likely trajectory
leading to the current point,
\begin{eqnarray}
{S_\mathrm{r}}({\bf \Gamma},t)
& = &
S_\mathrm{r}(\overline{ \underline {\bf \Gamma}})
+ S_\mathrm{r}(t) - \frac{1}{t} \int_0^t \mathrm{d}t'\, S_\mathrm{r}(t')
\nonumber \\ & \approx &
S_\mathrm{r}(\overline{ \underline {\bf \Gamma}}) .
\end{eqnarray}
The final two terms in the first equality
come from the conservation law for weight in non-equilibrium systems,
(see \S 8.3.1 of Ref.~[\onlinecite{NETDSM}]).
These invoke the total entropy, which at time $t'$ is
$S_\mathrm{r}(t') = k_\mathrm{B} \ln \int \mathrm{d} {\bf \Gamma}'
e^{S_\mathrm{r}({\bf \Gamma}',t') /k_\mathrm{B} } $.
Since this is independent of ${\bf \Gamma}$,
the final two terms of the first equality
may be neglected in the point entropy,
the final equality.
This is equivalent to incorporating them into the partition function
of the non-equilibrium probability.

Invoking the reduction condition
the reservoir entropy for a point in the sub-system phase space
is formally
\begin{eqnarray} \label{Eq:SrGt}
S_\mathrm{r}({\bf \Gamma},t)
& = &
S_\mathrm{st}({\bf \Gamma},t) + S_\mathrm{dyn}({\bf \Gamma},t)
 \\ \nonumber & \equiv &
S_\mathrm{st}({\bf \Gamma},t)
-
\int_{0}^t \mathrm{d} t'\,
\dot S_\mathrm{st}^0(\overline{\bf \Gamma}(t'|{\bf \Gamma},t) ,t') .
\end{eqnarray}
This result
holds for both mechanical and  thermodynamic non-equilibrium systems.

With this the formally exact expression for the phase space
probability of a non-equilibrium system is
\begin{equation} \label{Eq:wp(t)}
\wp({\bf \Gamma},t) =
\frac{1}{Z(t)} e^{S_\mathrm{r}({\bf \Gamma},t) /k_\mathrm{B}} .
\end{equation}
To be useful,
an explicit formula for the most likely trajectory is required,
and this is derived below.

\section{Fluctuation Forms} \label{Sec:Fluc-Form}

The purpose of this and the following \S \ref{Sec:NEsdEoM}
is to derive the stochastic, dissipative
equations of motion that correspond to the non-equilibrium probability.
In fact, the derivation could be bypassed
in favor of the physical interpretation  discussed in the conclusion.

\subsection{The Reservoir Entropy}

A complementary expression for the reservoir entropy
associated with each point in the sub-system phase space
may be obtained from fluctuation theory.
Let $\overline {\bf \Gamma}(t) = \left< {\bf \Gamma}(t) \right>$
be the most likely configuration of the sub-system at time $t$
and let $\bm{\gamma} \equiv {\bf \Gamma} - \overline {\bf \Gamma}(t)$
be the fluctuation or the departure from the most likely point.
\comment{ 
The static part of the reservoir entropy is generally known.
One example has already been given,
the canonical mechanical non-equilibrium system,
Eq.~(\ref{Eq:SreqMechWk}).
Another example is steady heat flow, which is the canonical
thermodynamic non-equilibrium system,
in which case\cite{AttardV,NETDSM}
\begin{equation}
S_\mathrm{st}({\bf \Gamma},t) =
\frac{- E_0({\bf \Gamma},t)}{T_0}
-\frac{E_1({\bf \Gamma},t)}{T_1} ,
\end{equation}
where $E_n({\bf \Gamma},t)$ is the $n$-th moment
of the sub-system energy,
$T_0$ is essentially the average temperature of the reservoirs,
and $T_1$ is essentially the temperature gradient due to the reservoirs.
} 
The non-equilibrium reservoir entropy is maximized
at $ {\bf \Gamma}(t) = \overline {\bf \Gamma}(t)$,
or, equivalently, $\bm{\gamma} = 0$,
and so in the expansion of it about this point the linear term vanishes,
\begin{equation} \label{Eq:SrGt-fluct}
S_\mathrm{r}({\bf \Gamma},t)
=
\overline{S}_\mathrm{r}(t)
+ \frac{1}{2} \overline {S''}_\mathrm{\!\!\!\! r}\,(t) : \bm{\gamma}\bm{\gamma}
+ {\cal O}( \bm{\gamma}^3) .
\end{equation}
Here $\overline{S}_\mathrm{r}(t)$ is the reservoir entropy most
likely produced to date, and the fluctuation matrix is
$ \overline {S''}_{\!\!\!\mathrm{r}}(t)  \equiv
\left.  \nabla \nabla S_\mathrm{r}( {\bf \Gamma},t)
\right|_{\overline{\bf \Gamma}(t)}$.

An expansion of the static part of the reservoir entropy about the
most likely configuration yields
\begin{equation}
S_\mathrm{st}({\bf \Gamma},t)
=
\overline S_\mathrm{st}(t)
+ \overline {S'_\mathrm{st}}(t) \cdot \bm{\gamma}
+  \frac{1}{2} \overline {S_\mathrm{st}''}(t)  : \bm{\gamma}\bm{\gamma}
+ {\cal O}( \bm{\gamma}^3).
\end{equation}
The gradient on the most likely trajectory is
$\overline {S'_\mathrm{st}}(t)
= \nabla S_\mathrm{st}(\overline{\bf \Gamma}(t),t)$.

In the dissipative equations of motion that are derived below
the gradient of the reservoir entropy appears.
In this one may approximate the full reservoir entropy fluctuation  matrix
by the static part of the entropy  fluctuation  matrix,
\begin{eqnarray} \label{Eq:olS''r=olS''st}
\nabla S_\mathrm{r}({\bf \Gamma},t)
& = &
\overline {S_\mathrm{r}''}(t)  \cdot \bm{\gamma}
+ {\cal O}( \bm{\gamma}^2)
\nonumber \\ & \approx &
\overline {S_\mathrm{st}''}(t)  \cdot  \bm{\gamma} .
\end{eqnarray}
There are three justifications for this approximation.
The first is  that the fluctuations about
the non-equilibrium state are determined by the current molecular
structure of the sub-system,
which is  the local equilibrium structure,
and they are therefore determined by the static part of the reservoir entropy.
The second is that the fluctuations about the most likely non-equilibrium state
have the same character and symmetries as equilibrium fluctuations.
In particular, the matrix representing them
must be block diagonal in the parity representation,
just as $\overline{S''}_\mathrm{\!\!\!\! st}(t)$ is.
The third justification follows from the resulting equations of
motion and is given in the conclusion.
This approximation has been found to be accurate
in computer simulation tests for both mechanical
\cite{AttardVIII,AttardIX,Attard09}
and thermodynamic
\cite{AttardV,AttardIX} non-equilibrium systems.

Since the gradient of the static part of the reservoir entropy
has the expansion,
\begin{eqnarray}   \label{Eq:DSreq}
\nabla S_\mathrm{st}({\bf \Gamma},t)
& = &
\overline {S'}_\mathrm{st}(t)
+ \overline {S_\mathrm{st}''}(t) \cdot \bm{\gamma}
+ {\cal O}( \bm{\gamma}^2) ,
\end{eqnarray}
one sees that the above approximation for the gradient of the reservoir entropy
implies that
\begin{eqnarray}  \label{Eq:grad-Sdyn}
\nabla S_\mathrm{dyn}({\bf \Gamma},t)
& = &
\nabla S_\mathrm{r}({\bf \Gamma},t)
-
\nabla S_\mathrm{st}({\bf \Gamma},t)
\nonumber \\ & \approx &
- \nabla S_\mathrm{st}(\overline{\bf \Gamma}(t),t)
+ {\cal O}( \bm{\gamma}^2) .
\end{eqnarray}
This is a time dependent constant in phase space.

\subsection{The Second Entropy} \label{Sec:NE-S2-fluct}

The second entropy is the entropy of transitions.
For the transition
$\{{\bf \Gamma}_1,t_1\} \rightarrow \{{\bf \Gamma}_2,t_2\}$,
it has general quadratic form
\begin{eqnarray} \label{Eq:S2-NE-1}
S^{(2)}({\bf \Gamma}_2,t_2;{\bf \Gamma}_1,t_1)
& = &
\frac{1}{2} A: \bm{\gamma}_2 \bm{\gamma}_2
+
B: \bm{\gamma}_2 \bm{\gamma}_1
+
\frac{1}{2} C: \bm{\gamma}_1 \bm{\gamma}_1
\nonumber \\ && \mbox{ }
+ \frac{1}{2} \left[
\overline S_\mathrm{r}(t_2) +  \overline S_\mathrm{r}(t_1) \right].
\end{eqnarray}
This is written in terms of the departures from the most likely value
and is the analogue of the fluctuation expression
for the first entropy given in the preceding sub-section.
The coefficients here are a function of the two times,
and will be written $A(t_{21},t)$, $B(t_{21},t)$, and $C(t_{21},t)$,
where $t \equiv (t_2+t_1)/2$
and $t_{21} \equiv t_2 - t_1 = -t_{12}$.
Usually $t$ will not be shown explicitly.
The final time dependent constant term
arises from the reduction theorem:
for time dependent weights,
the transition weight is normalized to the geometric mean
of the two terminal states, \S 8.3.1 of Ref.~[\onlinecite{NETDSM}].

The coefficients are second derivative matrices
and therefore $A$ and $C$ are symmetric matrices.
Because $S^{(2)}({\bf \Gamma}_2,t_2;{\bf \Gamma}_1,t_1)
= S^{(2)}({\bf \Gamma}_1,t_1;{\bf \Gamma}_2,t_2)$,
one must have
\begin{equation} \label{Eq:S2A=Cne}
A(t_{21}) = C(t_{12}) \mbox{ and }  B(t_{21}) = B(t_{12})^\mathrm{T} .
\end{equation}
This may be termed the statistical symmetry requirement
and it simply reflects the usual rule of unconditional probability:
the probability of
${\bf \Gamma}_1$ at $t_1$ and ${\bf \Gamma}_2$ at $t_2$
is the same as
the probability of
${\bf \Gamma}_2$ at $t_2$ and ${\bf \Gamma}_1$ at $t_1$.

The small time expansions of the coefficients must be of the form
\begin{equation}
A(\tau,t) = \frac{-1}{|\tau|} \Lambda(t)^{-1} + A_0(t) + \hat \tau A_0'(t)
+ {\cal O}(\tau)
\end{equation}
and
\begin{equation}
B(\tau,t) = \frac{1}{|\tau|} \Lambda(t)^{-1} + B_0(t) + \hat \tau B_0'(t)
+ {\cal O}(\tau) ,
\end{equation}
with $\Lambda(t)$, $A_0(t)$, $A_0'(t)$, and $B_0(t)$ being symmetric,
and $B_0'(t)$ being antisymmetric.
Also $\Lambda(t)$ must be positive definite
and $\hat \tau \equiv \mbox{sign } \tau$.
The term proportional to $1/\tau$ can be shown to vanish.\cite{NETDSM}
The functional form of the expansion can best be justified by the final results.

Maximizing the second entropy with respect to $\bm{\gamma}_2$,
one obtains the most likely end point of the transition
\begin{eqnarray} \label{Eq:olgamma2}
\overline{\bm{\gamma}}_2
& = & -A(t_{21})^{-1} B(t_{21})  \cdot \bm{\gamma}_1
\nonumber \\  & = &
\bm{\gamma}_1
+ t_{21} \Lambda [ A_0' + B_0' ] \cdot\bm{\gamma}_1
+ | t_{21} | \Lambda [ A_0 + B_0 ] \cdot\bm{\gamma}_1
\nonumber \\   && \mbox{ }
+{\cal O}(t_{21}^2) .
\end{eqnarray}

This expression contains two terms:
a reversible term proportional to $t_{21} $,
and an irreversible term proportional to $|t_{21} |$.
The nomenclature comes from the behavior
of the reverse or backward transition from $ \{ {\bm{\gamma}}_2,t_2 \}$,
the end point of which can be labeled $ \{ {\bm{\gamma}}_3,t_1 \}$.
This is
\begin{eqnarray}
\overline{\bm{\gamma}}_3
& = &
\overline{\bm{\gamma}}_2
+ t_{12} \Lambda [ A_0' + B_0' ] \cdot \overline{\bm{\gamma}}_2
+ | t_{12} | \Lambda [ A_0 + B_0 ] \cdot \overline{\bm{\gamma}}_2
\nonumber \\  & = &
\bm{\gamma}_1
+ 2 | t_{12} | \Lambda [ A_0 + B_0 ] \cdot {\bm{\gamma}}_1
+{\cal O}(t_{21}^2).
\end{eqnarray}
One sees that the reversible term has canceled.
If it were not for the irreversible term,
the reverse transition would end up at the starting point
of the original transition.
It is the presence of the term proportional to the absolute value of the
time step that makes the transition irreversible.
This irreversible term is necessary for the second law of thermodynamics.

The most likely terminus inserted into the fluctuation expression
for the second entropy, Eq.~(\ref{Eq:S2-NE-1}),
must reduce it to the first entropy for the starting point,
$S({\bf \Gamma}_1,t_1)$,
plus half the change in entropy over the time interval.\cite{NETDSM}
Invoking the fluctuation form for the first entropy,
Eq.~(\ref{Eq:SrGt-fluct}),
one obtains
\begin{eqnarray}
\lefteqn{
S^{(2)}(\overline{\bf \Gamma}_2,t_2;{\bf \Gamma}_1,t_1)
} \nonumber \\
& = &
S({\bf \Gamma}_1,t_1)
+ \frac{1}{2} \left[ \overline S_\mathrm{r}(t_2)
- \overline S_\mathrm{r}(t_1) \right]
\nonumber \\ & = &
\frac{1}{2} \overline {S''_{\mathrm{r}}}(t_1) :
\bm{\gamma}_1 \bm{\gamma}_1
+ \frac{1}{2} \left[
\overline S_\mathrm{r}(t_1) + \overline S_\mathrm{r}(t_2) \right].
\end{eqnarray}
The final constant term here
is equal to that in Eq.~(\ref{Eq:S2-NE-1}), as required.
Equating the coefficients of the quadratic term on both sides
one obtains
\begin{equation} \label{Eq:S2red2}
C(t_{21},t)
- B(t_{21},t)^\mathrm{T} A(t_{21},t)^{-1} B(t_{21},t)
= \overline {S''_{\mathrm{r}}}(t_1) .
\end{equation}
This must hold for all  $t_{21}$.

Expanding the left-hand side  yields
\begin{eqnarray}
\mathrm{LHS} & = &
\frac{-1}{|t_{21}|} \Lambda^{-1} + A_0 - \hat t A_0'
+
\left[
\frac{1}{|t_{21}|} \Lambda^{-1} + B_0 - \hat t B_0' \right]
\nonumber \\ && \mbox{ } \times
\left\{ \mathrm{I}
+ t_{21} \Lambda [ A_0' + B_0' ]
+ | t_{21} | \Lambda [ A_0 + B_0 ] \right\}
\nonumber \\ &  & \mbox{ }
+{\cal O}(t_{21})
\nonumber \\ & = &
A_0 - \hat t A_0'
+ \hat t [ A_0' + B_0' ] + [ A_0 + B_0 ]
\nonumber \\ &  & \mbox{ }
+ B_0 - \hat t B_0'
+{\cal O}(t_{21})
\nonumber \\ & = &
2[ A_0(t) + B_0(t) ]
+{\cal O}(t_{21}).
\end{eqnarray}
The right-hand side is
\begin{equation}
\overline {S''_{\mathrm{r}}}(t_1) =
\overline {S''_{\mathrm{r}}}(t) - \frac{t_{21}}{2}
\frac{\mathrm{d}\overline {S''_{\mathrm{r}}}(t)}{\mathrm{d}t} .
\end{equation}
Hence
\begin{equation}
A_0(t) + B_0(t) =
\frac{1}{2} \overline {S''_{\mathrm{r}}}(t)   .
\end{equation}
With this,
the irreversible part of the conditionally most likely transition,
Eq.~(\ref{Eq:olgamma2}),
the part proportional to $|t_{21}|$,  is
\begin{eqnarray} \label{Eq:olRgamma}
\overline{\bf R}_\gamma({\bf \Gamma},t_{21},t)
& \equiv &
|t_{21}| \Lambda(t) \left[ A_0(t) + B_0(t) \right] \cdot \bm{\gamma}
\nonumber \\ & = &
\frac{|t_{21}|}{2} \Lambda(t) \overline {S''_{\mathrm{r}}}(t)
\cdot  \bm{\gamma}
\nonumber \\ & = &
\frac{|t_{21}|}{2} \Lambda(t) \cdot \nabla S_\mathrm{r}({\bf \Gamma},t) .
\end{eqnarray}

The reversible  part of the conditionally most likely transition,
Eq.~(\ref{Eq:olgamma2}),
the part  proportional to $t_{21}$,
must contain the adiabatic evolution.
This part may in addition contain
reversible contributions from the reservoir,
but since the interactions with the reservoir are only a perturbation
on the total interactions in the sub-system,
any reversible contributions are small
compared to the adiabatic contribution.
Neglecting these one has
\begin{equation} \label{Eq:dot-gam-0}
\Lambda [A_0' + B_0' ] \cdot \bm{\gamma}
= \dot{\bm{\gamma}}^0 ,
\end{equation}
where $\dot{\bm{\gamma}}^0 $ is the adiabatic velocity of the fluctuation.

There are two justifications for neglecting
any reversible reservoir contribution to the evolution of a fluctuation.
The primary reason for considering any reversible reservoir contribution
to be negligible is that the thermodynamic contribution
to a fluctuation should be time symmetric:
in the future and in the past the fluctuation is equally likely
to be closer to zero.
As far as the reservoir is concerned,
the regression of a fluctuation back to the most likely state
is as probable as
the progression of the fluctuation from the most likely state.
This means that the reservoir contribution to the transition
between fluctuation states should be an even function of time.

The secondary reason is the one mentioned above,
namely that the reservoir represents a perturbation on the sub-system,
and any  reversible reservoir contribution
would be dominated by the reversible adiabatic contribution.
Conversely,
it is necessary to retain the irreversible reservoir contribution
because it is the only such contribution to the evolution
and so it is not negligible relative to any other term
with the same time symmetry.

With these,
the conditionally most likely transition, Eq.~(\ref{Eq:olgamma2}),
becomes
\begin{eqnarray} \label{Eq:olgamma2-b}
\overline{\bm{\gamma}}_2
& = &
\bm{\gamma}_1
+ t_{21} \dot{\bm{\gamma}}^0
+ \overline{\bf R}_\gamma({\bf \Gamma},t_{21},t)
+{\cal O}(t_{21}^2)
 \\  \nonumber & = &
\bm{\gamma}_1
+ t_{21} \dot{\bm{\gamma}}^0
+ \frac{|t_{21}|}{2} \Lambda(t) \cdot \nabla S_\mathrm{r}({\bf \Gamma},t)
+{\cal O}(t_{21}^2) .
\end{eqnarray}
To the exhibited order,
${\bf \Gamma}$ can be replaced by ${\bf \Gamma}_1$ or ${\bf \Gamma}_2$,
and $t$ can be replaced by $t_1$ or $t_2$.
The physical interpretation of this is
that the evolution of the fluctuation is the sum of a
reversible adiabatic term due to the internal interactions
within the sub-system,
and an irreversible term that arises from the reservoir.
This latter term depends upon the gradient in the entropy
and it represents the thermodynamic driving force
toward the most likely trajectory $\overline{\bf \Gamma}(t)$.

Recalling that the fluctuation is $ \bm{\gamma}(t)
\equiv {\bf \Gamma} - \overline{\bf \Gamma}(t)$,
this may be rearranged  to give the conditionally  most likely
point in phase space itself,
\begin{eqnarray} \label{Eq:olG2}
\lefteqn{
\overline{\bf \Gamma}(t_2|{\bf \Gamma}_1,t_1)
} \nonumber \\
& = &
{\bf \Gamma}_1
+ t_{21} \dot{\bf \Gamma}^0(t)
+ \frac{|t_{21}|}{2} \Lambda(t) \cdot \nabla S_\mathrm{r}({\bf \Gamma},t)
\nonumber \\ & & \mbox{ }
+ \overline{\bf \Gamma}(t_2)
- \overline{\bf \Gamma}(t_1)
-  t_{21} \dot{\overline{\bf \Gamma}}\,\!^0(t)
\nonumber \\
& = &
{\bf \Gamma}_1
+ t_{21} \dot{\bf \Gamma}^0(t)
+ \frac{|t_{21}|}{2} \Lambda(t) \cdot \nabla S_\mathrm{st}({\bf \Gamma},t)
\nonumber \\ & & \mbox{ }
+  \frac{t_{21} - |t_{21}|}{2} \Lambda(t) \cdot
\nabla S_\mathrm{st}(\overline{\bf \Gamma}(t),t) .
\end{eqnarray}
In obtaining the final equality,
the approximation
$ \overline {S''_{\mathrm{r}}}(t) \approx  \overline {S''_{\mathrm{st}}}(t)$
has been used for the reservoir gradient,
Eqs~(\ref{Eq:olS''r=olS''st}), (\ref{Eq:DSreq}), and (\ref{Eq:grad-Sdyn}).
Further,
by evaluating the first three terms at $\overline{\bf \Gamma}(t_1)$
in the forward direction,
the reservoir contribution to the evolution of the most likely trajectory
has been identified as
\begin{equation}
\overline{\bf \Gamma}(t_2)
- \overline{\bf \Gamma}(t_1)
-  t_{21} \dot{\overline{\bf \Gamma}}\,\!^0(t)
=
\frac{t_{21}}{2} \Lambda(t) \cdot
\nabla S_\mathrm{st}(\overline{\bf \Gamma}(t),t) .
\end{equation}
(See the physical interpretation in the conclusion.)
Note that this evolution is not purely adiabatic,
$ \dot{\overline{\bf \Gamma}}(t)
\ne \dot{\overline{\bf \Gamma}}\,\!^0(t)$,
which is to say that it contains contributions from the reservoir.
Even though the entropy gradient vanishes on the most likely trajectory,
$\nabla S_\mathrm{r}(\overline{\bf \Gamma}(t),t) = 0$,
it is the static part of the gradient,
$\nabla S_\mathrm{st}(\overline{\bf \Gamma}(t),t) \ne 0$,
that drives the non-adiabatic evolution.
Also,
because the most likely trajectory is a single valued function of time,
its evolution has to be reversible,
and hence this reservoir contribution
is proportional to $t_{21}$ rather than to $|t_{21}|$.
One sees that this last expression is the same as the penultimate expression
evaluated at $ {\bf \Gamma}_1  =  \overline{\bf \Gamma}(t_1)$.

The reservoir contributions to the trajectory evolution
may be gathered together and one can write
\begin{equation}
\overline{\bf \Gamma}_2(t_2|{\bf \Gamma}_1,t_1)
=
{\bf \Gamma}_1
+ t_{21} \dot{\bf \Gamma}^0({\bf \Gamma},t)
+ \overline{\bf R}({\bf \Gamma},t_{21},t)
+{\cal O}(t_{21}^2).
\end{equation}
Here the most likely reservoir `force' is
\begin{eqnarray} \label{Eq:olR}
\lefteqn{
\overline {\bf R}({\bf \Gamma},t_{21},t)
}  \\ \nonumber
& \equiv &
\frac{|t_{21}|}{2} \Lambda(t) \cdot \nabla S_\mathrm{st}({\bf \Gamma},t)
+  \frac{t_{21} - |t_{21}|}{2} \Lambda(t) \cdot
\nabla S_\mathrm{st}(\overline{\bf \Gamma}(t),t)
\nonumber \\ & = &
\left\{
\begin{array}{ll} \displaystyle
\frac{|t_{21}|}{2} \Lambda(t) \cdot \nabla S_\mathrm{st}({\bf \Gamma},t)
, & t_2 > t_1 \\
\displaystyle
\frac{|t_{21}|}{2} \Lambda(t) \cdot
\left[ \nabla S_\mathrm{st}({\bf \Gamma},t)
- 2 \nabla S_\mathrm{st}(\overline{\bf \Gamma}(t),t)
\right]
, & t_2 < t_1 .
\end{array} \right.
\nonumber
\end{eqnarray}
One sees from this that on a forward trajectory, $ t_2 > t_1 $,
the most likely trajectory $\overline{\bf \Gamma}(t)$,
which is \emph{a priori} difficult to calculate,
is not required.
Only the static part of the reservoir entropy is needed,
an expression for which is always known and relatively easy to evaluate.
On a backward trajectory, $ t_2 < t_1 $,
$\overline{\bf \Gamma}(t)$ is required,
and this makes the calculation of $S_\mathrm{dyn}({\bf \Gamma},t)$
a challenge. This is addressed in \S \ref{Sec:AdTraj} below.

The physical interpretation of the terms in the equation of motion
are discussed in the conclusion.


\section{Stochastic, Dissipative Equations of Motion} \label{Sec:NEsdEoM}

The total reservoir force is the sum of the above dissipative
term and a stochastic contribution,
${\bf R} = \overline {\bf R} + \tilde {\bf R}$.
The origin of the random force may be seen by
rearranging the second entropy explicitly into  Gaussian form,
which is to say a fluctuation about  the conditionally most likely terminus,
$\overline{\bm \gamma}_2$.
Since the remainder reduces to the first entropy this is
\begin{eqnarray} \label{Eq:S2g2g1}
S^\mathrm{(2)}({\bf \Gamma}_2,t_2;{\bf \Gamma}_1,t_1)
& = &
\frac{- \Lambda^{-1}}{2|t_{21}|}
: \left[\bm{\gamma}_2-\overline{\bm \gamma}_2\right]^2
+ S_\mathrm{r}({\bf \Gamma}_1,t_1)
\nonumber \\ &  & \mbox{ }
+  \frac{1}{2} \left[ \overline  S_\mathrm{r}(t_2)
- \overline  S_\mathrm{r}(t_1) \right] .
\end{eqnarray}
Because terms linear in $t_{21}$ have been neglected
in the expansions of the coefficients,
this expression for the second entropy neglects terms
${\cal O}(\bm{\gamma}^2t_{21})$.

Since the transition probability is the exponential of the second entropy
divided by Boltzmann's constant,
one sees from this that the evolution in phase space
has a stochastic character,
$\tilde {\bf R} \equiv \bm{\gamma}_2-\overline{\bm \gamma}_2
= {\bf \Gamma}_2 - \overline{\bf \Gamma}(t_2|{\bf \Gamma}_1,t_1) $.
The random force has zero mean,
$\langle \tilde{\bf R} \rangle = 0 $,
and  it is is Gaussian distributed with variance
\begin{equation} \label{Eq:<RR>}
\left< \tilde{\bf R}(t) \, \tilde{\bf R}(t) \right>
=  |t_{21}| k_\mathrm{B} \Lambda(t) .
\end{equation}
Random forces at different time steps are uncorrelated.

The corresponding stochastic, dissipative evolution equation is
\begin{eqnarray} \label{Eq:olGamma2}
{\bf \Gamma}_2 & = &
\overline{\bf \Gamma}(t_2|{\bf \Gamma}_1,t_1)
+ \tilde{\bf R}(t_{21},t)
\nonumber \\ & = &
{\bf \Gamma}_1
+ t_{21} \dot{\bf \Gamma}\!\,^0({\bf \Gamma}_1,t)
+ \overline {\bf R}({\bf \Gamma},t_{21},t)
\nonumber \\   && \mbox{ }
+ \tilde{\bf R}(t_{21},t)
+{\cal O}(t_{21}^2) .
\end{eqnarray}

The relationship between the variance of the stochastic force
and the coefficient of the dissipative force
may be called the fluctuation dissipation
theorem for the non-equilibrium system.
It shows that the magnitude of the fluctuations
is linearly proportional to the magnitude of the dissipation,
both being determined by $\Lambda$,
the leading-order coefficient of the second entropy expansion.
In  consequence, one can never
have dissipation without fluctuations.
The dissipative term $\overline {\bf R}$ is the thermodynamic driving force
that involves the gradient in entropy.
It is called this because dissipation
is the rate of entropy production,
which is the velocity times the gradient in the entropy.
One also sees from this and the stochastic dissipative evolution
equation that the variance of the random force is proportional
to the magnitude of the time step, $|t_{21}|$,
which makes it an irreversible contribution to the evolution.

The phase space point may be divided into its position and
momentum components,
${\bf \Gamma} = \{{\bf \Gamma}_\mathrm{q}, {\bf \Gamma}_\mathrm{p} \}$.
Apart from the factor of mass,
the change in the position component over a time step
is just the momentum component times the time step,
at least to first order in the time step,
$ {\bf \Gamma}_\mathrm{q}(t+\Delta_t)
= {\bf \Gamma}_\mathrm{q}(t) + \Delta_t {\bf \Gamma}_\mathrm{p}
+ {\cal O}(\Delta_t^2)$.
This is of course just the position component of the adiabatic evolution.
The stochastic, dissipative contributions from the reservoir
only enter the position evolution at second order in the time step,
${\bf R}_\mathrm{q}({\bf \Gamma},\Delta_t,t) \sim {\cal O}(\Delta_t^2)$.
Consequently,
the stochastic, dissipative equations of motion
in component form are
\begin{eqnarray} \label{Eq:SDEoMne}
{\bf \Gamma}_\mathrm{2q}
& = &
{\bf \Gamma}_\mathrm{1q}
+ t_{21} \dot {\bf \Gamma}^0_\mathrm{q}({\bf \Gamma},t)
+{\cal O}(t_{21}^2) ,
\\ \nonumber
{\bf \Gamma}_\mathrm{2p}
& = &
{\bf \Gamma}_\mathrm{1p}
+ t_{21} \dot {\bf \Gamma}^0_\mathrm{p}({\bf \Gamma},t)
+  {\bf R}_\mathrm{p}({\bf \Gamma},t_{21},t)
+{\cal O}(t_{21}^2) .
\end{eqnarray}
The adiabatic contributions are just Hamilton's equations,
\begin{equation}
\dot {\bf \Gamma}^0_\mathrm{q}
=
\frac{\partial{\cal H}({\bf \Gamma},t)}{\partial {\bf \Gamma}_\mathrm{p} }
, \mbox{ and }
\dot {\bf \Gamma}^0_\mathrm{p}
=
\frac{-\partial{\cal H}({\bf \Gamma},t)}{\partial {\bf \Gamma}_\mathrm{q} } .
\end{equation}
Typically, and most simply,
one may take the fluctuation matrix to be diagonal,
$ \Lambda(t) = \lambda  \mathrm{I}_\mathrm{pp}$.

\section{Adiabatic Trajectory} \label{Sec:AdTraj}

There are fundamentally two distinct computational approaches
where the above analysis is both necessary and useful
for non-equilibrium systems.
The first is stochastic molecular dynamics
based upon the stochastic, dissipative equations of motion,
Eq.~(\ref{Eq:SDEoMne}).
Since these are calculated forward in time,
Eq.~(\ref{Eq:olR}) shows that only the
relatively trivial static part
of the reservoir entropy is required.
Algorithms and results for stochastic molecular dynamics
for non-equilibrium systems have been given.\cite{AttardIX}

The second computational approach is non-equilibrium Monte Carlo
simulation,
and for this the phase space probability, Eq.~(\ref{Eq:wp(t)}),
and hence the reservoir entropy, Eq.~(\ref{Eq:SrGt}),
are required.
The dynamic part of the latter involves an
integral over the backwards most likely trajectory,
$\overline{\bf \Gamma}(t'|{\bf \Gamma},t)$, $t' \le t$,
which, from Eq.~(\ref{Eq:olR}), requires
the gradient of the static part of the reservoir entropy
at $\overline{\bf \Gamma}(t')$.
Since an expression for the latter is not explicitly available,
an approach has been tested wherein
the backwards most likely trajectory has been replaced
by the backwards adiabatic trajectory.
Computer simulations for steady heat flow\cite{AttardV}
and for a driven Brownian particle\cite{Attard09}
have shown this replacement to be accurate
and the resultant non-equilibrium Monte Carlo simulation algorithm
to be computationally feasible.
In this section this approximation is given and justified.

As has been mentioned,
the crucial distinction between an equilibrium and a non-equilibrium system
is that the probability distribution for the latter
depends upon the sign of the molecular velocities,
$ \wp({\bf \Gamma},t) \ne \wp({\bf \Gamma}^\dag,t)$.
For a sub-system phase space point
${\bf \Gamma} = \{{\bf q}^N, {\bf p}^N\}$,
the conjugate phase space point is the
one with the velocities reversed,
${\bf \Gamma}^\dag = \{{\bf q}^N, (-{\bf p})^N\}$.
Since the static part of the reservoir entropy
is a purely equilibrium quantity, it necessarily has even parity,
$S_\mathrm{st}({\bf \Gamma},t)
= S_\mathrm{st}({\bf \Gamma}^\dag,t) $.
This of course means that the dynamic part of the reservoir entropy
cannot be even,
$S_\mathrm{dyn}({\bf \Gamma},t) \ne S_\mathrm{dyn}({\bf \Gamma}^\dag,t) $.
Further, because the non-equilibrium aspects of the system
are a perturbation on the equilibrium aspects,
one can neglect the even projection of
$S_\mathrm{dyn}({\bf \Gamma},t)$
in comparison with $S_\mathrm{st}({\bf \Gamma},t)$ and so write
\begin{equation} \label{Eq:Sr=Sst+Sdynodd}
S_\mathrm{r}({\bf \Gamma},t)
\approx
S_\mathrm{st}({\bf \Gamma},t)
+ S_\mathrm{dyn}^\mathrm{odd}({\bf \Gamma},t) .
\end{equation}
The odd projection of the dynamic part of the reservoir entropy is
\begin{eqnarray} \label{Eq:dotSreq0odd}
S_\mathrm{dyn}^\mathrm{odd}({\bf \Gamma},t)
& \equiv &
\frac{1}{2} \left[
S_\mathrm{dyn}({\bf \Gamma},t)
-
S_\mathrm{dyn}({\bf \Gamma}^\dag,t)
\right]
\nonumber \\ & = &
\frac{-1}{2}
\int_{0}^t \mathrm{d} t'\,
\left[
\dot S_\mathrm{st}^0(\overline{\bf \Gamma}(t'|{\bf \Gamma},t) ,t')
\right. \nonumber \\ && \left. \mbox{ }
-
\dot S_\mathrm{st}^0(\overline{\bf \Gamma}(t'|{\bf \Gamma}^\dag,t) ,t')
\right] .
\end{eqnarray}

\begin{figure}[t!]
\centerline{
\resizebox{8cm}{!}{\includegraphics{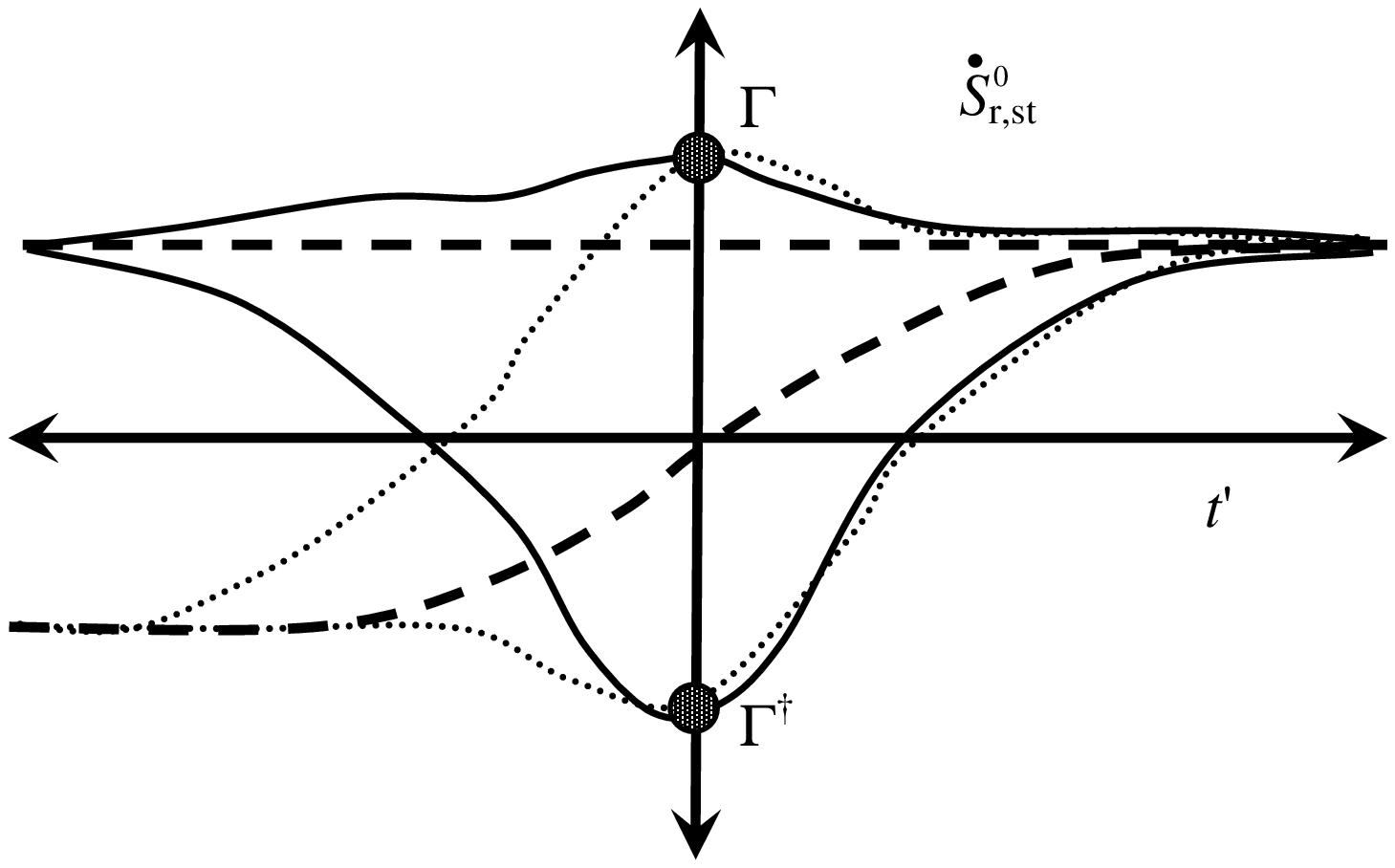}} }
\caption{\label{Fig:Trajs}  
Sketch of $\dot S_\mathrm{st}^0$
for a steady state non-equilibrium  system,
in the future and in the past,
starting from a likely phase space point ${\bf \Gamma}$,
and from its unlikely conjugate ${\bf \Gamma}^\dag$.
The solid curves are on most likely trajectories,
$\overline {\bf \Gamma}(t'|{\bf \Gamma},0)$
and $\overline {\bf \Gamma}(t'|{\bf \Gamma}^\dag,0)$,
and the dotted curves are on adiabatic trajectories,
${\bf \Gamma}^0(t'|{\bf \Gamma},0)$
and ${\bf \Gamma}^0(t'|{\bf \Gamma}^\dag,0)$.
The dashed line and curve are the respective asymptotes.
From Ref.~[\onlinecite{NETDSM}].}
\end{figure}

The behavior of the adiabatic rate of entropy production
on various trajectories is sketched in Fig.~\ref{Fig:Trajs}.
On the most likely trajectory the asymptotes are
\begin{equation}
\dot S_\mathrm{st}^0(\overline{\bf \Gamma}
(t'|{\bf \Gamma},t),t')
\rightarrow
\overline{\dot S_\mathrm{st}^0}(t') , \;\; |t'-t| \rightarrow \infty .
\end{equation}
This asymptote arises from the fact that with overwhelming
probability the system came from its most likely value in the past
(and will return there in the future), independent of the current
phase space point of the sub-system.
For a mechanical non-equilibrium system,
the Hamiltonian has to be extended into the future,
$\tilde {\cal H}({\bf \Gamma},t';t)
= {\cal H}({\bf \Gamma},2t-t')$ for $t'>t$,
and this causes the future asymptote to depend upon the current time,
$\overline{\dot S_\mathrm{st}^0}(t';t)$.
This extended Hamiltonian is an even function of time
about the current time $t$.
For a steady state system,
the asymptote is independent of $t'$ and $t$.

In contrast,
the asymptotic behavior on the adiabatic trajectory is
\begin{eqnarray}
\dot S_\mathrm{st}^0({\bf \Gamma}^0
(t'|{\bf \Gamma},t),t')
& \sim &
\mbox{sign}(t'-t) \,
\overline{\dot S_\mathrm{st}^0}(t')  .
\end{eqnarray}
This holds for $|t'-t| \agt \tau_\mathrm{relax}$,
where $\tau_\mathrm{relax}$ is a relaxation time
that is long enough for the system to reach its asymptote,
but not so long that the structure has changed significantly,
$|\tau_\mathrm{relax} \dot S_\mathrm{st}^0 | \ll |S_\mathrm{st}|$.
(One does not need to impose this condition
for the  dissipative trajectory
because the interactions with the reservoir maintain
the structure of the sub-system.)
For an isolated system,
the structure represents a fluctuation,
and $\dot S_\mathrm{st}^0 $ represents its regression,
which must be an odd function of time,
at least for a steady state system.
For $t'> 0$, the adiabatic asymptote
and the actual asymptote approximately coincide,
which is just Onsager's regression hypothesis.\cite{Onsager31}

In view of the trajectories shown in Fig.~\ref{Fig:Trajs}
and the above discussion,
the odd projection of the dynamic part of the reservoir entropy
may be transformed from an integral over the most likely trajectory
to an integral over the adiabatic trajectories.
Successive transformations yield
\begin{eqnarray} \label{Eq:Srdynodd-ol=0}
\lefteqn{
-2 S_\mathrm{dyn}^\mathrm{odd}({\bf \Gamma},t)
}  \\
& = &
\int_{0}^t \mathrm{d}t'\,
\left[
\dot S_\mathrm{st}^0
(\overline{\bf \Gamma}(t'|{\bf \Gamma},t),t')
-
\dot S_\mathrm{st}^0(\overline{\bf \Gamma}
(t'|{\bf \Gamma}^\dag,t),t') \right]
\nonumber \\ & \approx &
\int_{t}^{2t} \mathrm{d}t'\,
\left[
\dot S_\mathrm{st}^0
(\overline{\bf \Gamma}(t'|{\bf \Gamma},t),t')
-
\dot S_\mathrm{st}^0(\overline{\bf \Gamma}
(t'|{\bf \Gamma}^\dag,t),t') \right]
\nonumber \\ & \approx &
\int_{t}^{2t} \! \mathrm{d}t' \!
\left[
\dot S_\mathrm{st}^0({\bf \Gamma}^0(t'|{\bf \Gamma},t),t')
-
\dot S_\mathrm{st}^0({\bf \Gamma}^0(t'|{\bf \Gamma}^\dag,t),t') \right]
\nonumber \\  & = &
\int_{0}^t \! \mathrm{d}t' \!
\left[
\dot S_\mathrm{st}^0({\bf \Gamma}^0(t'|{\bf \Gamma}^\dag,t)^\dag,t')
-
\dot S_\mathrm{st}^0({\bf \Gamma}^0(t'|{\bf \Gamma},t)^\dag,t') \right] .
\nonumber
\end{eqnarray}
The first equality is the area between the solid curves
in the left half of the figure.
The second equality is the area between the solid curves
in the right half of the figure.
This follows because the dissipation on
the most likely trajectory is to a good approximation even in time.
The third equality is the area between the dotted curves
in the right half of the figure.
This follows from Onsager's regression hypothesis.
The fourth equality is the area between the dotted curves
in the left half of the figure.
This follows from the time reversibility of Hamilton's equations of motion,
\begin{equation}
{\bf \Gamma}^0(t'|{\bf \Gamma}^\dag,t)
=
{\bf \Gamma}^0(2t-t'|{\bf \Gamma},t)^\dag .
\end{equation}
For a mechanical non-equilibrium system,
the extended system Hamiltonian preserves this property.

For a thermodynamic, steady state, non-equilibrium system,
the adiabatic rate of change of the static part of the reservoir entropy
is  $ \dot S_\mathrm{st}^0({\bf \Gamma})
= \dot {\bf \Gamma}^0 \cdot \nabla  S_\mathrm{st}({\bf \Gamma})$,
and this has odd parity, $ \dot S_\mathrm{st}^0({\bf \Gamma}^\dag)
= -\dot S_\mathrm{st}^0({\bf \Gamma})$.
(For  time varying, reservoir induced, thermodynamic gradients
(i.e.\ non steady state),
 $ \dot S_\mathrm{st}^0({\bf \Gamma},t)$ is of mixed parity.)
For a mechanical non-equilibrium system,
the adiabatic rate of change of the static part of the reservoir entropy is
$ \dot S_\mathrm{st}^0({\bf \Gamma},t)
= - T^{-1}\partial U^\mathrm{ext}({\bf \Gamma},t)/\partial t$,
and this has even parity, $ \dot S_\mathrm{st}^0({\bf \Gamma}^\dag)
= \dot S_\mathrm{st}^0({\bf \Gamma})$.
Accordingly one can take the conjugate of the trajectories
in the final equality
of the above expression for the dynamic part of the reservoir entropy
and define
\begin{eqnarray}
S_\mathrm{dyn}^\mathrm{odd;0}({\bf \Gamma},t)
& \equiv &
\frac{-1}{2}
\int_{0}^t \! \mathrm{d}t' \!
\left[\dot S_\mathrm{st}^0({\bf \Gamma}^0(t'|{\bf \Gamma},t),t')
\right. \nonumber \\ & & \left. \mbox{ }
-
\dot S_\mathrm{st}^0({\bf \Gamma}^0(t'|{\bf \Gamma}^\dag,t),t')
\right] .
\end{eqnarray}
In these two cases one has
\begin{equation} \label{Eq:olG->G0}
S_\mathrm{dyn}^\mathrm{odd}({\bf \Gamma},t)
= \pm S_\mathrm{dyn}^\mathrm{odd;0}({\bf \Gamma},t),
\end{equation}
with the positive sign applying to
a thermodynamic, steady state, non-equilibrium system,
and the  negative sign applying to
a mechanical non-equilibrium system.
The left hand side invokes backward most likely trajectories,
and the right hand side invokes  backward adiabatic trajectories.

It is obvious from the figure that the integrand asymptotes to zero.
This means that the lower limit of the integral can be replaced by
$t-\tau$ for some convenient interval $\tau >0$.
Although the integrand is an exact differential, there is no point
in analytically evaluating the integral because the actual value at
the lower limit would be required,
$S_\mathrm{dyn}^\mathrm{odd}({\bf \Gamma},t) =
\pm [ S_\mathrm{st}({\bf \Gamma}^0(t-\tau|{\bf \Gamma},t),t-\tau)
- S_\mathrm{st}({\bf \Gamma}^0(t-\tau|{\bf \Gamma}^\dag,t),t-\tau) ]/2$.
(Although $\dot S_\mathrm{st}^0$ has the same asymptote
starting at ${\bf \Gamma}$ and at ${\bf \Gamma}^\dag$,
there is a finite difference between the respective asymptotes of
$ S_\mathrm{st}$ that corresponds to the area between
the two curves in the left half of the figure.)
It takes no more computational effort to perform the
quadrature numerically than it does to calculate the adiabatic
trajectories backward to their lower limit.

In summary,
this section argues  that in some circumstances
the odd projection of the
dynamic part of the reservoir entropy
is either dominant or is all that is required.
Further it says that the odd projection
of the dynamic part of the reservoir entropy
may be evaluated on the past adiabatic trajectories.
With this result,
one does not need to evaluate the most likely backwards trajectory,
and hence one does not need $\overline {\bf R}({\bf \Gamma},\Delta_t,t)$,
for $\Delta_t < 0$.
This means that explicit knowledge of $\overline {\bf \Gamma}(t)$
is not required, which is a great advantage.
As mentioned at the beginning of this section,
this adiabatic expression for the dynamic part of the reservoir entropy
has been tested with computer simulations
of both mechanical and thermodynamic non-equilibrium systems
and it has been found to be accurate.

\subsubsection{Green-Kubo Relations}

The validity and utility of the expression for the non-equilibrium
probability will now be illustrated with
a simple derivation of the Green-Kubo relations.
\cite{Onsager31,Green54,Kubo66}
These relate
the hydrodynamic transport coefficients to
the equilibrium time correlation functions of the fluxes.

For the particular case of heat flow,
the static part of the reservoir entropy is\cite{NETDSM,AttardI}
\begin{equation} \label{Eq:Sst-E0-E1}
S_\mathrm{st}({\bf \Gamma})
= \frac{-E_0({\bf \Gamma})}{T_0} - \frac{E_1({\bf \Gamma})}{T_1} ,
\end{equation}
where the $n$th energy moment in the $z$-direction is
$E_n({\bf \Gamma}) \equiv \int \mathrm{d}{\bf r}\,
\epsilon({\bf r};{\bf \Gamma}) z^n$,
with $\epsilon({\bf r};{\bf \Gamma})$ being the energy density
at ${\bf r}$.
Also the zeroth temperature is the mid-temperature of the two reservoirs,
$T_0^{-1} \equiv [ T_+^{-1} + T_-^{-1} ]/2 = T^{-1} + {\cal O} (\nabla T)^2$,
and the first temperature is essentially the temperature gradient
imposed by them,
$T_1^{-1} \equiv [ T_+^{-1} - T_-^{-1} ]/L_z
= -  T^{-2}\nabla T  + {\cal O} (\nabla T)^2$.

The instantaneous heat flux,
a phase function of the isolated sub-system, is essentially
the adiabatic rate of change of the first energy moment,
\cite{Onsager31,NETDSM,AttardI}
\begin{equation} \label{Eq:JE=E1/V}
J_\mathrm{E}({\bf \Gamma})
\equiv
\dot E_1^0({\bf \Gamma}) /V,
\end{equation}
where $V$ is the volume of the sub-system.
Due to energy conservation of the isolated system,
$ \dot E_0^0({\bf \Gamma}) = 0 $
and $\dot S^0_\mathrm{st}({\bf \Gamma}) = - \dot E_1^0({\bf \Gamma})/T_1$.

Fourier's law gives the heat flow in the presence of an applied
thermal gradient, and it is
\cite{Kubo78,Zwanzig01,Bellac04,Pottier10,NETDSM}
\begin{equation}
\overline J_\mathrm{E} = - \lambda \nabla T  ,
\end{equation}
where $\lambda$ is the thermal conductivity.
The left hand side is the most likely heat flux,
which equals the average heat flux.
This law of course holds to linear order in the temperature gradient.

The average heat flux given by the present non-equilibrium theory is
\begin{eqnarray}
\lefteqn{
\left< J_\mathrm{E} \right>_\mathrm{non-equil}
} \nonumber \\
& = &
\frac{1}{V}
\int \mathrm{d} {\bf \Gamma} \;
\wp({\bf \Gamma}|T_0,T_1)   \dot E_1^0({\bf \Gamma})
\nonumber \\ & = &
\frac{1}{V}
\frac{\int \mathrm{d} {\bf \Gamma} \;
e^{[ S_\mathrm{st}({\bf \Gamma})+ S_\mathrm{dyn}({\bf \Gamma})]
/k_\mathrm{B}}  \dot E_1^0({\bf \Gamma})
}{
\int \mathrm{d} {\bf \Gamma} \;
e^{[ S_\mathrm{st}({\bf \Gamma})+ S_\mathrm{dyn}({\bf \Gamma})]
/k_\mathrm{B}} }
\nonumber \\ & = &
\frac{1}{Vk_\mathrm{B}}
\frac{
\int \mathrm{d} {\bf \Gamma} \;
e^{-E_0({\bf \Gamma}) /k_\mathrm{B}T_0}
\dot E_1^0({\bf \Gamma}) S_\mathrm{dyn}^\mathrm{odd;0}({\bf \Gamma})
}{
\int \mathrm{d} {\bf \Gamma} \;
e^{-E_0({\bf \Gamma}) /k_\mathrm{B}T_0} }
+ {\cal O}(\nabla T )^2
\nonumber \\ & = &
\frac{1}{2Vk_\mathrm{B}T_1}
\int_{-\tau}^0 \mathrm{d} t' \,
\left<
\dot E_1^0({\bf \Gamma})
\left[ \dot E_1^0({\bf \Gamma}^0(t'|{\bf \Gamma},0)
\right. \right.  \nonumber \\ && \left. \left.  \mbox{ }
- \dot E_1^0({\bf \Gamma}^0(t'|{\bf \Gamma}^\dag,0) \right]
\right>_\mathrm{equil}
\nonumber \\ & = &
\frac{- \nabla T }{2Vk_\mathrm{B}T_0^2}
\int_{-\tau}^\tau \mathrm{d} t' \,
\left<
\dot E_1^0({\bf \Gamma})
\dot E_1^0({\bf \Gamma}^0(t'|{\bf \Gamma},0)
\right>_\mathrm{equil} .
\end{eqnarray}
In the third equality
the exponentials have been expanded in powers of the temperature gradient
and second order terms have been neglected.
As well, terms that are the product of an even parity function,
$S_\mathrm{st}({\bf \Gamma})$
or $S_\mathrm{dyn}^\mathrm{even}({\bf \Gamma})$,
and an odd parity function, $\dot S_\mathrm{st}^0({\bf \Gamma})$
or $\dot E_1^0({\bf \Gamma})$,
vanish upon integration over phase space.
In addition, the most likely trajectory has been replaced
by the adiabatic trajectory, Eq.~(\ref{Eq:olG->G0}).
The equilibrium average arises because
$Z(T_0)^{-1} e^{-E_0({\bf \Gamma}) /k_\mathrm{B}T_0} $
is the Maxwell-Boltzmann distribution.
Comparing this to Fourier's law,
one can identify the thermal conductivity as
\begin{eqnarray}
\lambda
& = &
\frac{1}{2k_\mathrm{B}VT_0^2}
\int_{-\tau}^\tau \mathrm{d} t' \,
\left<
E_1^0({\bf \Gamma})
\dot E_1^0({\bf \Gamma}^0(t'|{\bf \Gamma},0)
\right>_\mathrm{equil}
\nonumber \\ & = &
\frac{1}{2Vk_\mathrm{B}T_0^2}
\left<
\dot E_1^0(t)
\left[ E_1(t+\tau) - E_1(t-\tau) \right] \right>_\mathrm{equil} .
\end{eqnarray}
The right hand side is independent of $\tau$
for for $\tau \agt \tau_\mathrm{relax}$.
This can be written in a number of different ways,
but all involve the equilibrium time correlation function
of the heat flux or an integral thereof.
This is a typical example of a Green-Kubo relation.
\cite{Onsager31,Green54,Kubo66}
It is to be noted that the time correlation function
in any Green-Kubo relation
\emph{always} invokes adiabatic trajectories.
\cite{Kubo78,Zwanzig01,Bellac04,Pottier10,NETDSM}

From this analysis one sees that
the general formula for obtaining the Green-Kubo relations is
\begin{eqnarray}
\lefteqn{
\left< \dot S_\mathrm{st}^0({\bf \Gamma}) \right>_\mathrm{non-equil}
} \nonumber \\
& = &
\frac{\pm 1}{k_\mathrm{B}}
\left<
\dot S_\mathrm{st}^0({\bf \Gamma})\,
 S_\mathrm{dyn}^\mathrm{odd;0}({\bf \Gamma}) \right>_\mathrm{equil} ,
\end{eqnarray}
with the plus sign for steady state thermodynamic systems,
and the minus sign for mechanical non-equilibrium systems.
The fact that the theory gives the Green-Kubo relations
should give one confidence both in the adiabatic transformation
of the dynamic part of the reservoir entropy,
 Eq.~(\ref{Eq:olG->G0}),
and in the general expression for the phase space probability
for non-equilibrium systems,  Eq.~(\ref{Eq:wp(t)}).
Of course the  Green-Kubo relations are a linear theory,
whereas the present expression for the phase space probability
for non-equilibrium systems applies in all circumstances,
linear and non-linear.


\section*{Conclusion}

The non-equilibrium phase space probability distribution,
Eq.~(\ref{Eq:wp(t)}),
\[ 
\wp({\bf \Gamma},t) =
\frac{1}{Z(t)} e^{S_\mathrm{r}({\bf \Gamma},t) /k_\mathrm{B}} ,
\] 
is simply a formal statement that probability is
the exponential of the total entropy,
which equals the reservoir entropy because
the points in the phase space of the sub-system have no internal entropy.
This is formally the same as in the equilibrium case.
The specifically non-equilibrium concept is that
the reservoir entropy consists of a static and a dynamic part,
Eq.~(\ref{Eq:SrGt}),
\begin{eqnarray}
S_\mathrm{r}({\bf \Gamma},t)
& = &
S_\mathrm{st}({\bf \Gamma},t) + S_\mathrm{dyn}({\bf \Gamma},t)
\nonumber \\ \nonumber & \equiv &
S_\mathrm{st}({\bf \Gamma},t)
-
\int_{0}^t \mathrm{d} t'\,
\dot S_\mathrm{st}^0(\overline{\bf \Gamma}(t'|{\bf \Gamma},t) ,t') .
\end{eqnarray}
The static part is the ordinary equilibrium expression
that is based on exchange of conserved variables with the reservoir.
In a non-equilibrium system the conservation laws may not hold,
(e.g.\ the energy may change due to a time-varying external potential,
or energy gradients may change by internal relaxation processes),
and so the dynamic part of the reservoir entropy subtracts
the adiabatic or internal change from   the total change.
This adiabatic change is path dependent,
but in the thermodynamic limit
it can be obtained by integrating over the most likely
trajectory leading to the current point and neglecting fluctuations
about this trajectory.
In \S \ref{Sec:AdTraj} a useful adiabatic approximation
to the trajectory used in the dynamic part of the entropy was given.

In \S\S \ref{Sec:Fluc-Form} and \ref{Sec:NEsdEoM}
the stochastic, dissipative equations of motion
that correspond to this non-equilibrium probability density were given.
For a single time step, the phase space transition
$\{ {\bf \Gamma}_1,t_1 \} \rightarrow \{ {\bf \Gamma}_2,t_2 \} $
was found to be governed by the stochastic, dissipative equation
(\ref{Eq:olGamma2}),
\begin{eqnarray}
{\bf \Gamma}_2
& = &
{\bf \Gamma}_1 + t_{21} \dot {\bf \Gamma}^0
+ \frac{|t_{21}|}{2} \Lambda(t) \cdot \nabla S_\mathrm{st}({\bf \Gamma},t)
\nonumber \\ \nonumber && \mbox{ }
+ \frac{t_{21}-|t_{21}|}{2} \Lambda(t) \cdot
\nabla S_\mathrm{st}(\overline{\bf \Gamma}(t),t)
+ \tilde{\bf R}(|t_{21}|,t) .
\end{eqnarray}
There are four transition terms on the right hand side.

The first term is the adiabatic velocity,
which is due to the internal interactions within the sub-system,
and which would occur if it were isolated.
These are of course reversible (proportional to $t_{21}$).

The second term represents some of the reservoir forces on the sub-system.
These are driven by the change in reservoir entropy
due to the sub-system--reservoir interactions. %
It is $S_\mathrm{st}$ not $S_\mathrm{r}$
that accounts for the change in entropy
due to the exchange of conserved variables
between the sub-system and the reservoir.
This exchange is of course
carried by such interactions.
Hence the gradient of the static part of the reservoir entropy
provides the thermodynamic driving force for the transition.
The drag coefficient $ \Lambda(t)$
represents the strength of the statistical coupling
between the sub-system and the reservoir.
The thermodynamic gradient is toward the state of higher entropy,
and as such it applies  both forward
and backward in time,
(i.e.\ it is irreversible,   $\propto |t_{21}|$).

The form of this second term arose  in the thermodynamic gradient
in Eq.~(\ref{Eq:olG2})
where the fluctuation matrix for the total reservoir entropy
was approximated by that of the static part alone,
Eq.~(\ref{Eq:olS''r=olS''st}),
\[ 
\overline {S''}_\mathrm{\!\!\!\! r}\,(t)
\approx
\overline{S''}_\mathrm{\!\!\!\! st}(t)
, \mbox{ and }
\overline{S''}_\mathrm{\!\!\!\! dyn}(t) \approx 0 .
\] 
To the two original  justifications for this approximation
(that fluctuations about
the non-equilibrium state are determined by the current molecular structure,
and that these fluctuations
have the same symmetries as equilibrium fluctuations)
may now be added a third:
it is essential that the thermodynamic driving force be
$\nabla S_\mathrm{st}({\bf \Gamma},t)$
not $\nabla S_\mathrm{r}({\bf \Gamma},t)$,
since the former gives the change in entropy specifically due
to interactions and exchange between the sub-system and the reservoir.

The third term is basically the second term
evaluated on the most likely trajectory, $\overline{\bf \Gamma}(t)$.
This term arises as a correction to the second term
because by definition the most likely trajectory is reversible
(i.e.\ a single valued function of time can be chosen).
One can see that on the most likely trajectory
the irreversible parts of the second and third terms cancel
with each other leaving the deterministic transition
on this trajectory fully reversible.
For a forward trajectory, $t_{21} > 0$,
the coefficient of this third term vanishes.

The fourth term is the stochastic term.
It arises from the fact that the sub-system phase space is a
projection of the total system phase space,
and so the evolution of a trajectory in the sub-system
is not uniquely determined by a point  in the sub-system,
which is the definition of randomness.
The random force has zero mean,
$\langle \tilde{\bf R} \rangle = 0 $.
In the text the fluctuation form of
the second entropy showed that it had variance, Eq.~(\ref{Eq:<RR>}),
\[ 
\left< \tilde{\bf R}(t) \, \tilde{\bf R}(t) \right>
=  |t_{21}| k_\mathrm{B} \Lambda(t) .
\] 
This is the fluctuation-dissipation theorem.
The  magnitude and functional form of the drag coefficient $ \Lambda(t)$
have little effect on the statistical results
provided that  the fluctuation-dissipation theorem is satisfied.
With it
the non-equilibrium probability
is functionally stationary
under the evolution governed by the stochastic,
dissipative equations of motion
(see \S 8.3.5 of Ref.~[\onlinecite{NETDSM}]).



\begin{thebibliography}{99}

\bibitem{Feynman98}
Feynman, R. P. (1998),
\emph{Statistical Mechanics: Statistical Mechanics: A Set Of Lectures},
(Advanced Books Classics, Westview Press, 2nd ed.).

\bibitem{Pathria72}
Pathria, R. K. (1972),
\emph{Statistical Mecahnics},
(Pergamon Press, Oxford).

\bibitem{McQuarrie00}
McQuarrie, D. A. (2000),
\emph{Statistical Mecahnics},
(University Science Books, Sausalito)

\bibitem{TDSM}
Attard, P. (2002),
\emph{Thermodynamics and Statistical Mechanics:
Equilibrium by Entropy Maximisation},
(Academic Press, London).


\bibitem{Kubo78}
Kubo, R., Toda, M.,  and Hashitsume, N. (1978),
\emph{Statistical Physics II. Non-equilibrium Statistical Mechanics},
(Springer-Verlag, Berlin).


\bibitem{Zwanzig01}
Zwanzig, R. (2001),
\emph{Non-equilibrium Statistical Mechanics},
(Oxford University Press, Oxford).


\bibitem{Bellac04}
Le Bellac, M., Mortessagne, F., and Batrouni, G. G., (2004),
\emph{Equilibrium and Non-equilibrium Statistical Thermodynamics},
(Cambridge University Press, Cambridge).

\bibitem{Pottier10}
Pottier, N. (2010),
\emph{Non-equilibrium Statistical Physics:
Linear Irreversible Processes},
(Oxford University Press, Oxford).


\bibitem{NETDSM}
Attard, P. (2012),
\emph{Non-Equilibrium Thermodynamics and Statistical Mechanics:
Foundations and Applications},
(Oxford University Press, Oxford).






\bibitem{Yamada67}
Yamada, T. and Kawasaki, K. (1967),
Progr.\ Theor.\ Phys.\ {\bf 38}, 1031.

\bibitem{Yamada75}
Yamada, T. and Kawasaki, K. (1975),
Prog.\ Theo.\ Phys.\ {\bf  53}, 111.


\bibitem{AttardV}
Attard, P. (2006),
J. Chem.\ Phys.\ {\bf 124}, 224103.



\bibitem{AttardVIII}
Attard, P. and Gray-Weale, A. (2008),
J. Chem.\ Phys.\  {\bf 128}, 114509.  


\bibitem{AttardIX}
Attard, P. (2009),
J. Chem.\ Phys.\ {\bf 130}, 194113. 


\bibitem{Attard09}
Attard, P. (2009),
Phys.\ Rev.\ E {\bf 80}, 041126.


\bibitem{QSM4}
Attard, P. (2014),
``Quantum Statistical Mechanics. IV. Non-Equilibrium Probability Operator'',
arXiv: \ldots


\bibitem{Onsager31}
Onsager, L. (1931),
Phys.\ Rev.\ {\bf 37}, 405, and {\bf 38}, 2265.


\bibitem{Green54}
Green, M. S. (1954),
J. Chem.\ Phys.\ {\bf 22}, 398.

\bibitem{Kubo66}
Kubo, R. (1966),
Rep.\ Progr.\ Phys.\ {\bf 29}, 255.


\bibitem{AttardI}
Attard, P. (2004),
J.\ Chem.\ Phys.\ {\bf 121}, 7076. 

\end{thebibliography}
\end{document}